\documentclass[12pt,preprint]{aastex}

\lefthead{}
\righthead{Neon Abundances in B-Stars of the Orion Association: Solving the Solar
Model Problem?}

\begin{document}

\title{Neon Abundances in B-Stars of the Orion Association: Solving the Solar
model Problem?}

\author{Katia Cunha}
\affil{National Optical Astronomy Observatory\footnote{On leave from
Observat\'orio Nacional - MCT; Rio de Janeiro, Brazil},
P.O. Box 26732, Tucson, AZ 85726 USA; kcunha@noao.edu}

\author{Ivan Hubeny}
\affil{Steward Observatory, University of Arizona, Tucson, AZ 85712, USA;
hubeny@aegis.as.arizona.edu}

\author{Thierry Lanz}
\affil{Department of Astronomy, University of Maryland, College Park, MD 20742 USA;
tlanz@umd.edu}

\begin{abstract}
We report on non-LTE Ne abundances for a sample of B-type stellar members of the Orion
Association.  The abundances were derived by means of non-LTE fully
metal-blanketed model atmospheres and extensive model atoms with updated
atomic data.  We find that these young stars have a very homogeneous
abundance of A(Ne) = 8.27 $\pm$ 0.05.  This abundance is higher by $\sim$ 0.4 dex
than currently adopted solar value, A(Ne)=7.84, which is derived from
lines produced in the corona and active regions.  The general agreement
between the abundances of C, N, and O derived for B stars with the solar
abundances of these elements derived from 3-D hydrodynamical models
atmospheres strongly suggests that the abundance patterns of the light
elements in the Sun and B stars are broadly similar.  If this hypothesis is
true, then the Ne abundance derived here is the same within the uncertainties as
the value required to reconcile solar models with helioseismological
observations.
\end{abstract}


\section{INTRODUCTION}
One important recent result from studies of stellar atmospheres and
chemical compositions in stars is the downward revision in the abundances of
carbon, nitrogen and oxygen in the solar photosphere, which was obtained from
the adoption of time-dependant, hydrodynamical and 3-Dimensional
model atmosphere calculations (Asplund 2005). These more realistic model atmosphere
calculations indicate that the solar abundances should be lower by
roughly 0.2-0.3 dex when compared to abundances derived from hydrostatic 1-D calculations.
It soon became apparent that the significantly lower abundances in the Sun resulted in severe
inconsistencies between the solar models and measurements
from helioseismology. 

Different possibilities were investigated in order
to try to reconcile the solar models with the seismological observations, such as
updating opacities (Bahcall et al. 2004), changing diffusion rates (Guzik et al. 2005), or 
significantly changing
the adopted solar abundances of key elements, such as neon. Allowing for
a larger neon abundance, in particular, was justified because the neon abundance in the Sun
can be considered to be more uncertain given that it is not measured from lines formed in 
the solar photosphere.
Antia \& Basu (2005) constructed envelope models of the Sun, allowed for different 
abundance mixtures and focused on the density profile, which is determined from helioseismology.
More complete calculations were presented in
Bahcall, Basu \& Serenelli (2005) who constructed solar models,
consisting of the atmospheres plus the interior, and concluded that an adopted
neon abundance A(Ne)= 8.29 $\pm$ 0.05, would suffice in order to bring the solar models 
and seismological observations into an acceptable agreement.
Independently, measurements of neon abundances in
a sample of chromospherically active cool stars 
by Drake \& Testa (2005) indirectly supported high Ne abundances
in solar type stars.

It has been a long standing puzzle that the C, N and O abundances obtained for B-stars,
which are young, were typically lower than the at-the-time generally accepted solar abundances 
from Anders \& Grevesse (1989). These were puzzling results because from our simplest
understanding of how the Galaxy chemically evolves, it is not expected that
young stars in the solar vicinity would be less enriched than the Sun, which is much older.
Moreover, the abundances obtained from Galactic H II regions were also lower than
the accepted solar and in rough agreement with the B-star results.
These inconsistencies between the abundances of young stars and H II regions, on one
hand, and the Sun on the other, are reconciled nowadays with the revised
solar abundances from 3-D models.
In this context, it is therefore important and timely to derive accurate 
neon abundances in the atmospheres of early-type stars.
In this study, we report on non-LTE (NLTE) neon abundance calculations for
a sample of B stars members of the Orion association. The Ne abundances
in young stars
can independently shed light on the issue related to the reference
Ne abundance in the Galaxy.

\section{OBSERVATIONS}

The target stars are OB main-sequence members of the different stellar subgroups of
the Orion association and drawn from the sample analyzed by
Cunha \& Lambert (1992, 1994). Eleven stars were observed with the
2.1m telescope at the McDonald Observatory at high resolution (R=55,000)
using the Sandiford echelle. The spectra were obtained on Oct 27, 1994
and these have 26 echelle orders covering the total spectral range between 5390 and 6680 \AA.
The spectra were reduced with IRAF data package following standard procedures.

\section{Non-LTE Abundance Calculations}

The stellar parameters for the sample stars were derived in Cunha \& Lambert
(1992).
Non-LTE model atmospheres were computed using the TLUSTY code (Hubeny 1988, 
Hubeny \& Lanz 1995).
The model calculations assumed a constant microturbulence of 2 km/s.
Preliminary models for this study were taken from an extensive grid of 
NLTE line-blanketed model atmospheres of B stars (Lanz \& Hubeny, in prep.).
The final models were computed for the actual effective temperatures
and surface gravities of our program stars, and adopting an extended Ne model
atom. The BSTAR model grid is analogous to our OSTAR2002 grid (Lanz \& Hubeny
2003), the only difference being the addition of lower ionization stages of the
most important species. Concretely, the following ions were considered explicitly
in the BSTAR grid models: H I--II, He I--III, C I--V, N I--VI, O I--VI, Ne I--V, 
Mg II--III, Al II--IV, Si II--V, S II--VI, and Fe II--VI. 

The Ne model atom constructed consists of 79 levels of Ne I, 138 levels of Ne II,
38 levels of Ne III, 12 levels of Ne IV, plus ground state of Ne V.
The energies of the levels were taken from the Opacity Project database
TOPBASE (Cunto et al. 1993), updated by the more accurate experimental
level energies from the Atomic and Spectroscopic Database at NIST 
(Martin et al. 1999) whenever available. 
The $gf$-values were taken from the same sources. 
However, since the $LS$-coupling, on which the Opacity Project
calculations are based, is rather inaccurate for Ne I, we have used results and
the procedure suggested by Seaton (1998) to transform the level energies (and
designations) from $LS$-coupling to the more appropriate $jK$-coupling.
Also, we used a model atom that treats explicitly the fine structure of multiplets.
The photoionization cross-sections were taken from TOPBASE, and the collisional
excitation rates were considered using the Van Regemorter formula, and for
collisional ionization using the Seaton formula -- for a synopsis of expressions,
see Hubeny (1988).

The detailed synthetic spectra were computed using the {\it Interactive Data
Language} (IDL) interface SYNPLOT (Hubeny, unpublished) to the spectrum
synthesis program SYNSPEC (Hubeny, Lanz, \& Jeffery 1995). 
The abundances were obtained from the best fits
between observed and synthetic spectra of 8 Ne I transitions:
$\lambda$6506.5, 6402.2, 6383.0, 6334.5, 6266.5, 6163.5, 6143.1 and 6096.2 \AA.
We have computed NLTE line profiles (using NLTE atomic level populations
of all atoms and ions computed by TLUSTY). When computing detailed
synthetic spectra we found that the best fits to the observations were 
obtained for a microturbulent
velocity around 5 km/s, although the abundance results were quite insensitive to
the microturbulence parameter.

The nature of NLTE effects in Ne I line formation was already discussed
by Auer \& Mihalas (1973), Dworetsky \& Budaj (2000), and, in particular,
by Sigut (1999). Our models do not offer anything fundamentally different
from these studies; only we are using significantly more extended model
atoms, more recent atomic data, and fully blanketed model atmospheres.
The nature of NLTE effects for above listed optical lines of Ne I was
already explained in the earlier studies, namely since the lower levels
of the optical lines are connected to the ground state of Ne I by resonance
lines that are located in far UV, they are essentially in detailed balance
with the ground state. The optical lines thus behave like classical lines
in a two-level atom, that is the lower level is somewhat overpopulated, while
the upper level is depopulated (because of an imbalance of the number of 
excitations compared to de-excitations caused by the photon escape through 
the boundary). The source function is thus lower than the Planck function, 
and the optical lines are consequently predicted stronger than in LTE.
This in turn means that the deduced NLTE abundances are expected to be
lower that the LTE ones.

Our results are in qualitative agreement with the earlier studies.
We compared our calculations with the two previous studies (Sigut 1999 and
Dworetsky \& Budaj 2000). For one target star, HD35299,
we measured equivalent widths for the Ne I lines and computed Ne
abundances for an assumed microturbulence $\xi$=5.0 km/s. A modest systematic difference
was obtained for HD35299: $\delta$Ne(This study - Sigut) = 0.1 dex, for an average
of all lines (eqws computed by Sigut, priv. comm.) 
Dworetsky \& Budaj (2000), as discussed in their paper, find good agreement 
with the results by Sigut (1999). Dworetsky \& Budaj have also used TLUSTY for
their study, and made their TLUSTY-compatible Ne I atomic data input 
available online. We tested their input data, and found that they inadvertently
set up parameters for the collisional excitation rates in such a way that they
were effectively set to 0 (which means they they in fact maximized possible
NLTE effects). When we artificially modified our data by setting
collisional excitation rates also to zero, we obtained an excellent agreement
with Dworetsky \& Budaj. This also demonstrates that the collisional excitation 
rates are not critical for basic features of Ne I NLTE line formation.

\subsection{Ne Abundance Results and Previously Derived O Abundances}

The studied Orion stars span a significant range in effective temperature, 
from $\sim$ 20000 K to 29000 K (Table~1). In Figure 1 (top panel) we show that
there is no significant trend of the Ne abundances with the stars' effective temperature,
indicating the absence of important systematic errors in this study. 
A comparison between the LTE and
NLTE abundance trend with $T_{\rm eff}$ is shown in the bottom panel of Figure 1.
There is a trend of the derived LTE abundances with $T_{\rm eff}$,
which again demonstrates the inadequacy of LTE for abundance determinations,
as already shown by earlier investigations. 
The LTE line profiles were
obtained by setting the Ne level populations to their LTE values, while
the atmospheric structure (temperature, electron density, etc.), as well
as level population of other species, were kept at their NLTE values.
We stress that such line profiles are different from truly LTE line
profiles (computed for a consistent LTE structure and LTE level populations
of all species), but our ``LTE'' line profiles best demonstrate NLTE
effects in determining the Ne abundance.
The fact that LTE
abundances are {\em larger} than NLTE ones is in fact quite comforting in the
context of Ne abundance determinations, because otherwise one may be worried
whether the deduced high Ne abundance is an artifact of some spurious
NLTE effect caused for instance by some inadequacy in atomic data. However,
since the NLTE abundance is {\em smaller}, this potential worry can be
ruled out.

The Ne abundances obtained for the targets (listed in Table 1) 
are quite homogeneous
and show a small scatter that can be completely explained in terms of
uncertainties in the abundance determinations. 
The average Ne abundance for all target stars is A(Ne)=8.27 $\pm$ 0.05.
The uncertainties in the Ne I abundances can be estimated from
the sum in quadrature of the abundance uncertainties due to errors in the 
adopted stellar parameters, microturbulence, continuum location, as well as,
atomic data. We estimate that our derived Ne abundances are accurate to within
roughly 0.1 dex. The data available to us do not allow for Ne II abundances
to be derived in this study as a consistency check. (We note that Kilian-Montenbruck
et al. (1994) found high LTE Ne II abundances for OB stars.)

It is the Ne/O ratio, and not only neon, that is obtained from abundance measurements in the 
solar corona 
and this ratio constitutes an important ingredient in the construction of solar models. Before
a comparison can be done with the Orion results for B stars, it is important to stress 
that the oxygen and neon abundances for the Orion targets were not derived homogeneously.
The methodology presented in this study to
derive neon abundances consisted of a full NLTE treatment, including
NLTE line formation and the computation of fully-blanketed model atmospheres in NLTE.
For oxygen, however, we adopt the previously published results from Cunha \& Lambert (1994)
which were derived in LTE and using Kurucz LTE model atmospheres, and finally corrected by means 
of the NLTE calculations by  Becker \& Butler (1988). 
It is important then to verify   
whether the previously derived oxygen abundances for Orion are consistent with
results from the more sophisticated calculations presented here for neon.

As a consistency check on the published oxygen results, we re-derived the oxygen 
abundance for one of the Orion stars, HD35299, using TLUSTY/SYNSPEC.
We adopted the same published  
equivalent widths for O II lines and same stellar parameters but used the 
TLUSTY NLTE model atmospheres calculated for the Ne analysis. We calculated
O II abundances versus the microturbulence parameter and derived $\xi$= 5 km/s
and an oxygen abundance A(O)= 8.65 $\pm$ 0.05. This abundance compares favorably
with the oxygen abundance obtained in Cunha \& Lambert (1992), within the uncertainties.
Such agreement justifies the adoption of the published oxygen results in order
to investigate Ne/O ratios.

\section{DISCUSSION}

Solar photospheric abundances can be readily compared to 
meteoritic C1 chondrite abundances and good agreement is found for most of the elements,
or more specifically, for those elements that form rocks (see e.g. Lodders 2003).
Noble gases, as well as carbon, nitrogen and oxygen are volatiles and
their abundances are therefore significantly depleted in meteorites. For C,N, and O
one can rely on abundances measured in the solar photospheres, available 
from both 1-D and 3-D model atmospheres calculations.
For neon, however, the solar abundances are subject to further uncertainties due to
the absence of photospheric lines because even the lower excited states
of the Ne atom have very high energy. Alternatively, Ne abundances in the Sun are inferred
from measurements of Ne/O in the solar coronal gas, solar wind and solar energetic particles.
The most recent assessment of the Ne abundance in the Sun is obtained from
measurements of Ne/O in the solar corona and from energetic particles is A(Ne)=7.84 
(Asplund, Grevesse \& Noels 2005). The solar value according to Lodders (2003)
is just slightly higher (A(Ne)=7.87).

The neon abundances derived here for a sample of early-type stars in the Orion
association are found to be quite homogeneous. 
The average neon abundance for the studied stars (A(Ne)=8.27 $\pm$ 0.05) is higher than the 
quoted solar value by $\sim$ 0.4 dex.
In Figure 2 we show our Ne results versus oxygen abundances. 
The average oxygen abundance for the sample Orion stars is: A(O)=8.70 $\pm$ 0.09.
which is entirely consistent with a single oxygen abundance and agrees with
the solar abundance of A(O)=8.66 (Asplund et al. 2005).

Recent results from detailed calculations of oxygen abundances in a sample of 3 B0.5V 
stars in the Orion nebula indicate an average oxygen abundance of 
A(O)=8.63 $\pm$ 0.03 (Simon-Diaz et al. 2006).
The oxygen abundance data presented here is a subsample of the stars 
analyzed in Cunha \& Lambert (1994). For the full sample with 18 stars this previous study 
obtained an oxygen
abundance spread which was larger: A(O)= 8.72 $\pm$ 0.13, but marginally within the
abundance uncertainties. 
Our target stars in this study, however, span a tighter
range in oxygen abundances (A(O)= 8.70 $\pm$ 0.09) and are considered to represent 
a single oxygen abundance for all purposes.
Therefore, the B stars in the Orion association can be 
represented by a Ne/O ratio of 0.38, which is much higher than the currently
adopted solar value of 0.15 (Asplund et al. 2005).

The Orion Nebula has been the most extensively studied galactic H II region
and is recognized as the standard reference for nebular abundance studies in the Galaxy. 
Recently, Esteban et al. (2004) conducted a careful emission line study of several elements 
in the Orion nebula and obtained
A(O)=8.65 and A(Ne)=8.05 (or, Ne/O=0.25); these are gas 
abundances and corrections for any element trapped in grains have not been considered.
This nebular result is lower than our average for the Orion stars but higher
than the solar value (see Figure 2). 

\section{Conclusions}

Measurements of neon abundances in a variety of objects that can help define the
uncertain neon abundance in the Sun are potentially of great importance for solar physics.
We find the Ne abundance in B stars members of the Orion association is
significantly higher than the solar value by roughly 0.4 dex ($\sim$ 2.5X). 
We argue that the Ne abundances measured in young OB stars 
should be a good representation of the solar chemical composition, as is indicated
from the good agreement between the abundances in B stars and Sun for other elements 
such as C, N, and O. The high Ne abundances obtained here 
come to the rescue of the solar models that require, according to Bahcall, Basu \& Serenelli (2005), 
an increase in the Ne abundance by $\sim$2.8 $\pm$ 0.4. 

\acknowledgements
We especially thank C. Allende Prieto for discussions and for communicating with T. Sigut.
We also thank D. Arnett, L. Stanghellini, S. Wolff and D. Garnett for several discussions.
We thank V. V. Smith for helping with the observations.
The work reported here is supported in part by the National Science
Foundation through AST03-07534, AST03-07532, and NASA through NAG5-9213.

\clearpage

\begin{deluxetable}{lcccc}
\tablecaption{Sample Stars and Abundances}
\tablewidth{0pt}
\tablehead{
\colhead{Star} & \colhead{Teff} & \colhead{Log g} & \colhead{A(Ne)} & \colhead{A(O)}}
\startdata
HD35039 & 20550 & 3.74 & 8.25 & 8.60 \\
HD35299 & 24000 & 4.25 & 8.30 & 8.57 \\
HD35912 & 19590 & 4.20 & 8.21 & 8.70 \\
HD36285 & 21930 & 4.40 & 8.29 & 8.80 \\
HD36351 & 21950 & 4.16 & 8.24 & 8.76 \\
HD37356 & 22370 & 4.13 & 8.33 & 8.67 \\
HD37209 & 24050 & 4.13 & 8.28 & 8.83 \\
HD37744 & 24480 & 4.40 & 8.35 & 8.63 \\
HD36959 & 24890 & 4.41 & 8.21 & 8.76 \\
HD36591$^{a}$ & 26330 & 4.21 & 8.26 & 8.60 \\
HD36960 & 28920 & 4.33 & 8.21 & 8.72

\enddata
\tablecomments{(a): Recent results from IUE flux,
2MASS and Johnson magnitudes (Nieva \& Przybilla 2006)
indicate quite good agreement with the adopted
stellar parameters.
}
\end{deluxetable}

\clearpage

\begin{figure}
\epsscale{.8}\plotone{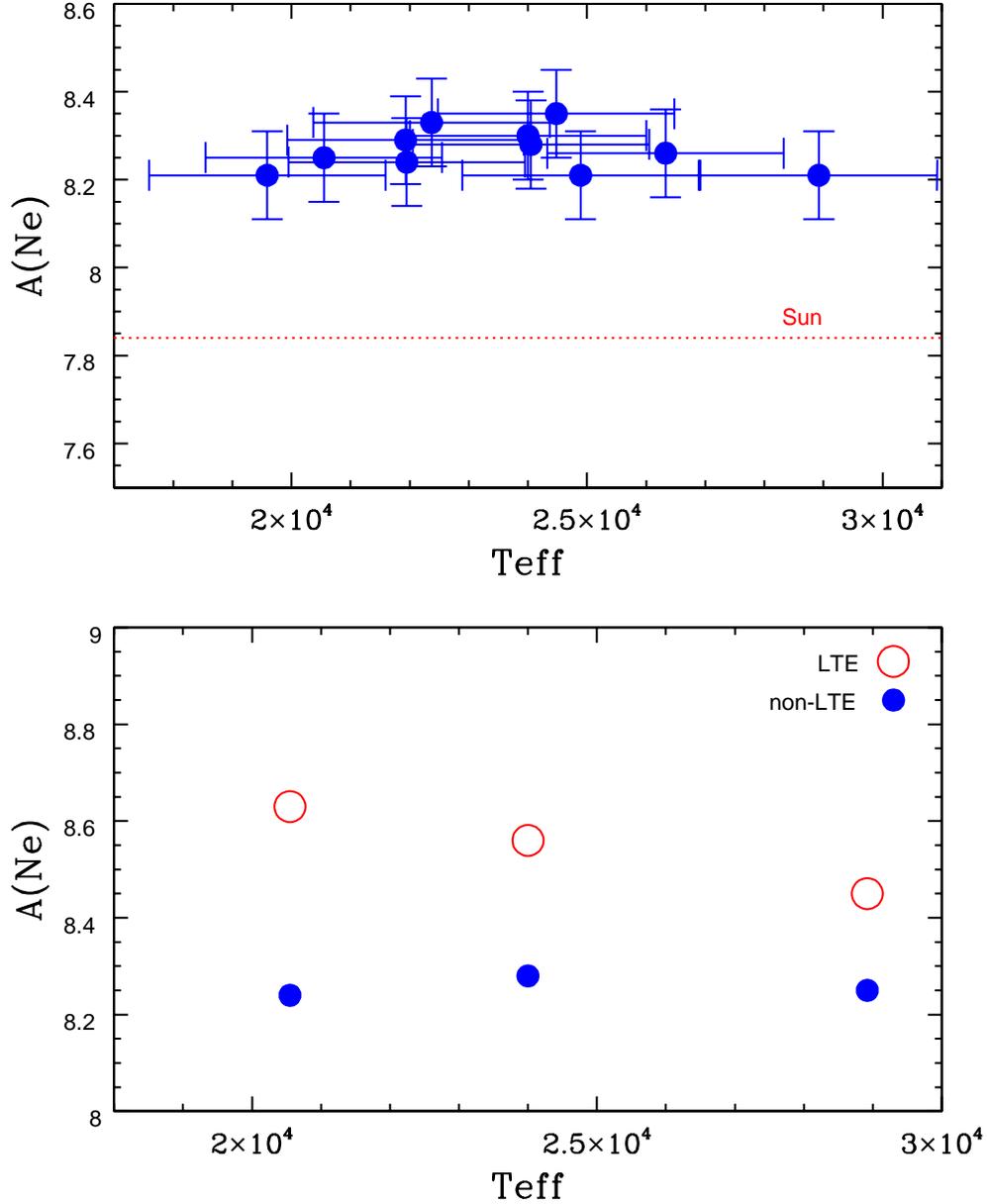}
\caption{\label{fig1} Top panel:
The Ne derived abundances versus the adopted
effective temperatures for the sample Orion stars. The $T_{\rm eff}$ range covered
here is relatively large and no trends are found for the derived neon abundances.
For comparison, we also indicate the currently adopted solar Ne abundance 
from Asplund et al. (2005) as the dashed line.
Bottom panel:
LTE (open circles) and NLTE (filled circles) abundances calculated for
the strongest Ne I line $\lambda$ 6402 \AA. This is the only line that
is strong enough to be measured in the hottest stars in our sample.
The LTE Ne abundances are
found to have a trend with effective temperature. This trend
is erased with the NLTE calculations.}
\end{figure}

\begin{figure}
\includegraphics[scale=0.3,angle=270]{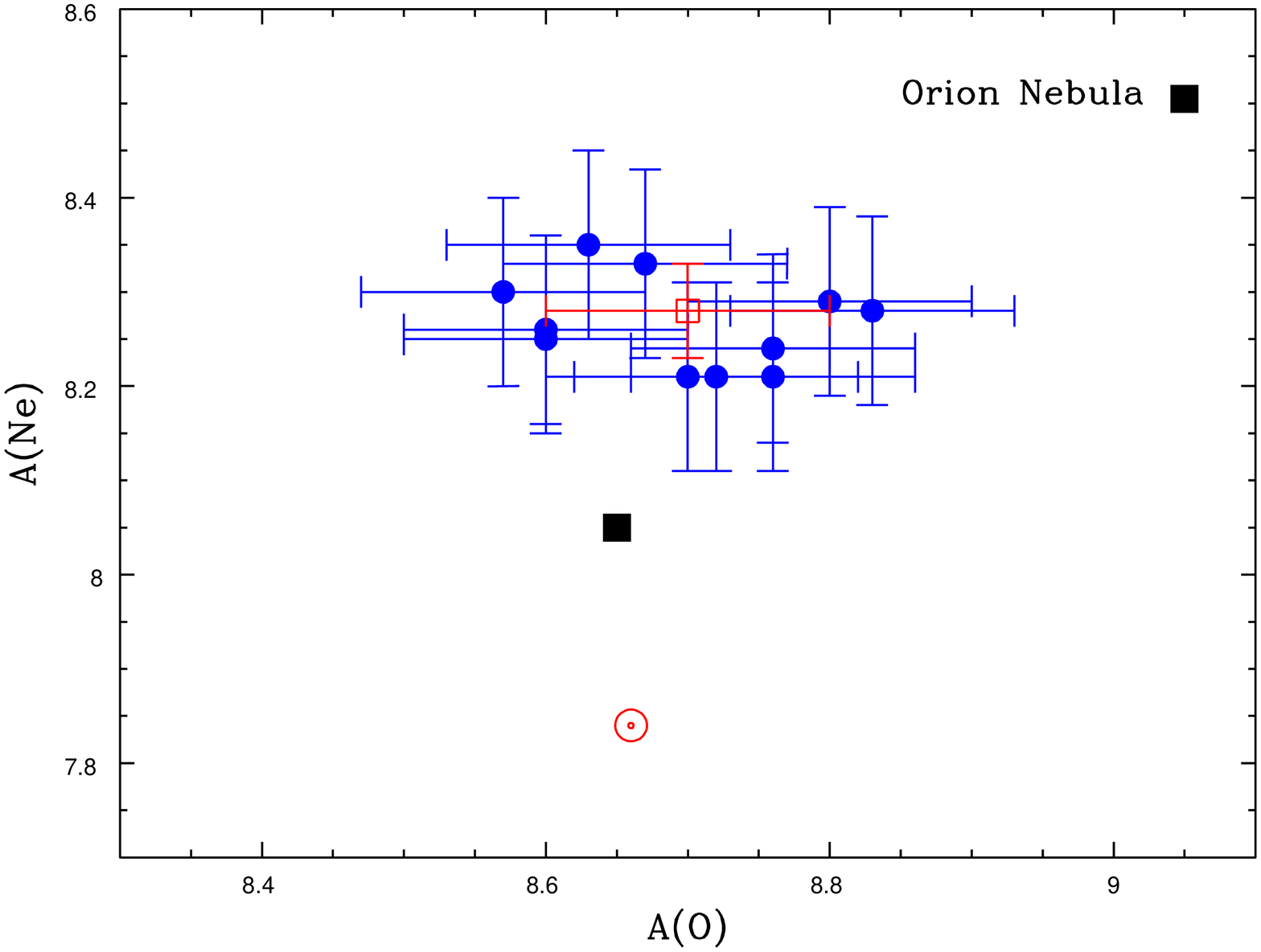}
\caption{\label{fig2} Neon abundances derived for the Orion B stars in this study versus the
oxygen abundances from Cunha \& Lambert (1994; filled circles). The average Ne and O
abundance for the studied sample is represented by the open square. For comparison
we also show the currently adopted solar value (Asplund et al. 2005), 
as well as the H II region abundance obtained for the
Orion nebula by Esteban et al. (2004; filled square), without accounting for any depletion
onto grains.}

\end{figure}


\begin{thebibliography}{}

\bibitem[]{336} Anders, E. \& Grevesse, N. 1989, Geochimica et Cosmochimica Acta, vol. 53, p. 197

\bibitem[]{338} Antia, M., \& Basu, S. 2005, ApJ, 621, L85

\bibitem[]{340} Asplund, M. 2005, Ann. Rev. Astron. Astrophys., 43, 481

\bibitem[]{342} Asplund, M., Grevesse, N., \& Sauval, A. J.2005 In Cosmic 
Abundances as Records of Stellar Evolution and Nucleosynthesis, ed. F. N. Bash, \&
T. G. Barnes p. 25

\bibitem[]{346} Auer, L. H., \& Mihalas, D. 1973, ApJ, 184, 151

\bibitem[]{348} Bahcall, J. N., Basu, S., \& Serenelli, A. M. 2005, ApJ, 631, 1281

\bibitem[]{350} Bahcall, J. N., Serenelli, A. M.,  Basu, S. 2005b, ApJ, 621, L85

\bibitem[]{352} Becker, S. R. \& Butler, K. 1988, A\&A, 201, 232 

\bibitem[]{354} Cunha, K. \& Lambert, D.L. 1992, ApJ, 399, 586

\bibitem[]{356} Cunha, K. \& Lambert, D.L. 1994, ApJ, 426, 170 

\bibitem[]{358} Cunto, W., Mendoza, C., Ochsenbein, F., \& Zeipen, C. J. 1993,
A\&A, 275, L5

\bibitem[]{361} Drake, J. J., \& Testa, P. 2005, Nature, 43, 525

\bibitem[]{363} Dworetsky, M. M., \& Budaj, J. 2000, MNRAS, 318, 1264 

\bibitem[]{365} Guzik, J. A., Watson, L. W., \& Cox, A. N. 2005, ApJ, 627, 1049

\bibitem[]{367} Grevesse, N., \& Sauval, A. J. 1998, Space Sci. Rev. 107, 665

\bibitem[]{369} Hubeny, I. 1988, Comput. Phys. Commun., 52, 103

\bibitem[]{371} Hubeny, I., \& Lanz, T. 1995, ApJ, 439, 875

\bibitem[]{373} Hubeny, I., Lanz, T., \& Jeffery, C. S. 1994, SYNSPEC - A User's
Guide, in Newsletter on Analysis of Astronomical Spectra No 20, St. Andrews Univ.

\bibitem[]{375} Kilian-Montenbruck, J., Gehren, T., \& Nissen, P. E. 1994, A\&A, 291, 757

\bibitem[]{376} Lanz, T., \& Hubeny, I. 2003, ApJS, 146, 417

\bibitem[]{378} Lester, J. B, Gray, R. O., \& Kurucz, R. L. 1986, ApJS, 61, 509

\bibitem[]{380} Martin, W. C., Sugar, J., Musgrove, A., Wiese, W. L., \& Fuhr, J. R.
1999, NIST Atomic Spectra Database

\bibitem[]{381} Nieva, M.F., \& Przybilla, N. 2006, ApJ, 639, L39

\bibitem[]{383} Seaton, M. J. 1998, J. Phys. B, 31, 5315

\bibitem[]{385} Sigut, T. A. A. 1999, ApJ, 519, 313

\bibitem[]{387} Simon-Diaz, S., Herrero, A., Esteban, C., \& Najarro, F. 2006, A\&A, 351, 366

\end{thebibliography}
\end{document}